\documentclass[aps,pre,twocolumn,amsmath,amssymb,superscriptaddress]{revtex4-1}

\usepackage{bm}
\usepackage{graphicx}
\usepackage{color,graphics}

\begin{document}

\title{How large can the electron to proton mass ratio be in Particle-In-Cell simulations of unstable systems?}

\author{A Bret}

\affiliation{ETSI Industriales, Universidad de Castilla-La Mancha, 13071
Ciudad Real, Spain,\\
Instituto de Investigaciones Energ\'{e}ticas y Aplicaciones Industriales,
13071 Ciudad Real, Spain.}

\author{M.E. Dieckmann}

\affiliation{VITA, Department of Science and Technology (ITN), Link\"{o}ping
University, 60174, Norrk\"{o}ping, Sweden}

\begin{abstract}
Particle-in-cell (PIC) simulations are widely used as a tool to investigate instabilities that develop between a collisionless plasma and beams of charged particles. However, even on contemporary supercomputers, it is not always possible to resolve the ion dynamics in more than one spatial dimension with such simulations. The ion mass is thus reduced below 1836 electron masses, which can affect the plasma dynamics during the initial exponential growth phase of the instability and during the subsequent nonlinear saturation. The goal of this article is to assess how far the electron to ion mass ratio can be increased, without changing qualitatively the physics. It is first demonstrated that there can be no exact similarity law, which balances a change of the mass ratio with that of another plasma parameter, leaving the physics unchanged.  Restricting then the analysis to the linear phase, a criterion allowing to define a maximum ratio is explicated in terms of the hierarchy of the linear unstable modes. The criterion is applied to the case of a relativistic electron beam crossing an unmagnetized electron-ion plasma.
\end{abstract}

\maketitle


\maketitle

\section{Introduction}
The dynamics of collision-less plasma far from its equilibrium is frequently examined with particle-in-cell (PIC) simulations. The unique capability of PIC codes to model such systems from first principles on macroscopic scales implies that they can bridge the gap between theory and experiment. For example, just a few years ago it was still unclear if relativistic shocks exist. It was not known whether the motion energy could be dissipated rapidly enough to sustain the shock discontinuity \cite{Brainerd2000,Waxman2006}. Such shocks have not yet been observed directly, because they do not exist in solar system plasma. Recent PIC simulations could shed light on how they develop in response to the filamentation (Weibel) instability \cite{SilvaApJ,Spitkovsky2008,Chang2008}. The particle acceleration and the generation of electromagnetic radiation within the context of active galactic nuclei \cite{GALLANT1992,FrederiksenApJ2004,DieckmannMNRAS2006,Spitkovsky2008a}, supernova remnants \cite{HOSHINO1992,DieckmannApJ2009} or gamma ray bursts \cite{Jaroschek2004,Jaroschek2005,MedvedevApJ2005,DieckmannMNRAS2006a} have also been investigated. PIC simulations are now instrumental in investigating the plasma thermalization within solar flares \cite{KarlickyApJ,Karlicky2009} and the dynamics of magnetic reconnection \cite{CheYoon,Lapenta}. On a completely different length and density scale, the fast ignition scenario for inertial confinement fusion \cite{Tabak} has prompted within the last decades many numerical works focusing on the propagation of charged particle beams in a collisionless plasma \cite{Pukhov96,Ren2004}.

Current simulations employ billions of computational particles, placing physically realistic PIC simulations within our reach \cite{MedvedevSpit2009}. The inclusion of ions in PIC simulations nevertheless remains a formidable challenge. As long as the system under scrutiny involves only electrons and positrons with the mass $m_e$, the time scale that must be resolved is typically the inverse electronic plasma frequency $\omega_e^{-1}\propto \sqrt{m_e}$. Running the simulation for hundreds or thousands $\omega_e^{-1}$ captures the evolution of the system way beyond its linear phase. Mobile protons or ions in the simulation result in an additional and much longer timescale. A PIC simulation must then resolve many inverse proton plasma frequencies $\omega_p^{-1}\propto \sqrt{M_p}$ and cover a time interval that is $F_p = \sqrt{M_p/m_e}\sim 42$ times longer, if the plasma is unmagnetized and if protons are the only ion species. A further penalty is introduced by the larger spatial scales of the ion structures. The size of the ion filaments is, for example, comparable to the ion skin depth $c/\omega_p$, while that of the electron filaments is $\sim c/\omega_e$. In principle, the simulation box size that is necessary to model electron-proton plasmas increases compared to that required by leptonic plasmas by a factor $\sim F_p^D$, where $D$ is the number of resolved spatial dimensions.

For this reason and until now, PIC simulations that use the correct electron-to-proton mass ratio are restricted to 1D systems \cite{DieckmannMNRAS2006}   and to 2D simulations that resolve a limited spatio-temporal domain \cite{HondaPRL}, while 2D PIC simulations that cover a large domain with regard to the ion scales or even 3D PIC simulations normally resort to reduced ion masses between 10-100 electron masses \cite{Spitkovsky2008,Niemiec2008,Martins2009}. Ion masses of up to 1000 electron masses have been used in a 2.5D simulation \cite{Spitkovsky2008} thanks to a low number of particles per cell, which is beneficial for the scalability of a domain-decomposed PIC code.
  Multidimensional plasma simulations that employ the correct mass ratio and capture the largest ion scales are possible, if the electron dynamics does not have to be resolved accurately. Implicit PIC schemes can dissipate away the energy contained in the smallest scales in a form of Landau damping \cite{Lapenta2006,Noguchi2007}. The cell size can then be increased beyond the plasma Debye length, without restricting the physical accuracy of the large-scale dynamics. However, if both the electron and the ion dynamics must be resolved simultaneously, the implicit PIC codes are equally costly as the explicit ones. The speeding up of the plasma evolution through a reduced ion mass will remain a necessity in particular for 3D simulations. It is thus important to assess how the plasma dynamics changes with this parameter.

  Various studies exist that demonstrate the importance of the electron-to-ion mass ratio for the plasma dynamics in several types of plasma processes. Parametric studies of plasma shocks have addressed this issue both in the non-relativistic \cite{Scholer2003,Scholer2004,Shimada2005} and relativistic \cite{Spitkovsky2008} regimes. Simulation studies of the interplay between electron phase space holes with the ions and its dependence on the mass ratio can be found in \cite{califanoIOns2007,EliassonPRL2004}. The impact of the mass ratio on the reconnection of magnetic field lines and the associated particle acceleration has been investigated in Ref. \cite{Ricci2004,GuoandLu2007}. However, these studies related to the effects of a reduced ion mass focus primarily on the nonlinear evolution of the simulation.

This article is a first systematic study of the consequences of a reduced ion mass within a theoretical framework. The impact of the mass ratio on the nonlinear coupling of the plasma dynamics across the different scales is not considered here; it is too complex and multifaceted. An example would be the enlargement of the foreshock of a perpendicular shock with an increasing ion mass, which influences the resulting instabilities and the thermalization of the shock-reflected ion beam \cite{Scholer2003,Scholer2004,Shimada2005,Lee2004}. We study here the spectrum of linearly unstable waves, which should depend on the mass ratio between the ions and the electrons. Ions only \emph{one} time ``heavier'' than electrons are obviously too light as they behave like positrons, not like protons. Is it therefore possible to draw a line, from which ions will start being ``too light'' to represent protons?

Even before answering this question, one could ask whether the PIC simulation plasmas could be governed by some similarity laws involving the mass ratio. Similarity theory has been applied successfully in hydrodynamics. It allows us to predict certain properties of an object from experiments performed with its miniaturised model. Well-known cases of such experiments involve pumps, turbines or aircrafts \cite{BrennenPumps,Birkhoff}. Similarity laws have also been derived for magnetic confinement fusion (see \cite{petty2008}, and references therein) or relativistic laser-plasma interactions and laboratory astrophysics \cite{Drury2000,Ryutov2001,gordienko2005}. Similarity laws would allow us to compensate a reduced mass ratio with some other parameter, by which the computational efficiency can be altered.

We show in section 2 that it is not possible to derive an universal description of the growth rate spectrum, which is not explicitly dependent on the mass ratio. Section 3 will therefore aim at providing a restricted solution to the problem. PIC simulations typically probe the long term nonlinear evolution of an \emph{unstable} beam-plasma system. The unstable spectrum usually contains more that one unstable mode, and these modes grow at different exponential rates. A hierarchy of unstable modes can be established in terms of their growth rate, and a criterion can be imposed on the mass ratio by demanding that this hierarchy be preserved. The consequences of this condition are then explicitly calculated for the case of a cold relativistic electron beam passing through a un-magnetized and cold plasma \cite{califano3}. Section 4 is the discussion, which brings forward a possible explanation why the shock formation in Ref. \cite{Spitkovsky2008} does apparently not depend on the mass ratio.

\section{Absence of a similarity law}
Consider a problem that involves $n$ independent dimensional variables $(x_1,\ldots,x_n)$, and $m$ fundamental dimensions such as meter, second, etc. The so-called Buckingham's method \cite{Buckingham} to reduce the number of variables and to derive similarity laws is the following.

Identify the pairs of $x_i$'s that share the same physical unit. If this is the case for the variables $x_{k_1}$ and $x_{k_2}$, then replace $(x_{k_1},x_{k_2})$ by $(x_{k_1},x_{k_1}/x_{k_2})$ in the list of variables. After iterating this process for any such pairs, we are left with the modified set of variables $(x_1,\ldots,x_m,x_{k_1}/x_{k_2},\ldots,x_{k_{l-1}}/x_{k_l})$ where $l+m=n$. Buckingham's ``$\Pi$ theorem'' then states that any unknown function of the form
\begin{equation}\label{eq:buck1}
f(x_1,\ldots,x_m,x_{k_1}/x_{k_2},\ldots,x_{k_{l-1}}/x_{k_l})=0,
\end{equation}
can indeed be cast under the form,
\begin{equation}\label{eq:buck2}
\phi(\pi_1,\ldots,\pi_{m-p},x_{k_1}/x_{k_2},\ldots,x_{k_{l-1}}/x_{k_l})=0,
\end{equation}
where the variables $\pi_i$ are dimensionless products of the initial $x_i$'s, and $p$ the number of fundamental dimensions among $(x_1,\ldots,x_m)$.

Consider as a simple illustration a swinging pendulum with the mass $M$ (kg) and the length $l$ (m), which oscillates with the constant period $T$ (s) in a gravitational field $g$ (m/s$^2$). The first step of Buckingham's method, namely the pairing of variables that share the same dimension, can be skipped here since all 4 variables $(M,l,T,g)$ have different dimensions (kg, m, s, m/s$^2$). We thus have here $m=4$ (4 variables) and only $p=3$ (kg, m, s) as $g$ does not add any extra fundamental dimension to the problem.

Buckingham's theorem states in this case that any function $f(M,l,T,g)=0$ can be expressed as $\phi(\pi_{m-p=1})=0$, so that the problem is eventually a function of \emph{one} single dimensionless parameter. A subsequent dimensional analysis shows that the universal parameter must be a power of $gT^2/L$. Clearly, the mass $M$ cannot participate in the dimensionless parameter, because no other variables could cancel its physical unit. The period $T$ is thus independent of the mass $M$ and only a function of the ratio $g/L$. Note that the theorem does not distinguish between ``input'' variables (what is known, e.g. $M,l,g$) and ``output'' variables (what is looked for, e.g. $T$). Each quantity is treated in the same way and all contribute to $m$ and to $p$.

Turning now to the present problem, we see from the first step of Buckingham's method that a similarity law without an explicit dependence on the mass ratio cannot exist. Whatever the list of variables describing the problem may be, the mass parameters $(m_e,M_p)$ will be a part of it. The first step of the process will just replace $(m_e,M_p)$ by $(m_e,m_e/M_p)$, and Buckingham's theorem  reduces the number of variables left once all the dimensionless trivial ratios have been formed. At any rate, Buckingham's theorem states that the mass ratio remains as an explicit parameter in the final reduced set of parameters.

There is therefore no hope of unraveling similarity laws connecting two systems (A) and (B) with different mass ratios. A change of the mass ratio cannot be ``compensated'' by a shift of the other variables. Buckingham's analysis of the problem proves an intuitively simple reasoning: electrons and ions define different time scales in terms of their respective mass. The time evolution of the system can be normalized to any one of them, but it cannot fit both at the same time. Note that although the rest of the article focuses on the linear regime of an electron beam plasma system, the present conclusion is very general, and valid for the overall evolution of any kind of system comprised of two species.

Let us initially assume that the ions are immobile, yielding an electron-to-ion mass ratio of zero. Electrons are therefore the only population bringing a mass into the parameter list. The electron mass $m_e$ will therefore appear among the $x_1\ldots x_m$ in Eq. (\ref{eq:buck1}). Buckingham's theorem here states that these $m$ dimensional variables can be replaced by $m-p$ dimensionless variables $\pi_1\ldots \pi_{m-p}$. Because no mass ratio can appear among the $l-1$ ratios which are arguments of the function $\phi$ in Eq. (\ref{eq:buck2}), the underlying equations can not rely explicitly on the electron mass. Equation (\ref{eq:buck2}) shows that $m_e$ must have been ``absorbed'' by one of the dimensionless $\pi$'s. This is precisely what is observed when dealing with such questions: the time parameter is frequently normalized to the electronic plasma frequency which includes the electron mass.

\section{Preserving the growth rate hierarchy of the unstable modes}
In ultra-relativistic laser-plasma interaction, similarity theory states that laser-plasma interactions with different $a_0=eA_0/m_ec^2$ and $n_e/n_c$ are similar, as soon as the similarity parameter $S=n_e /a_0n_c$ is the same \cite{gordienko2005} ($A_0$ is the laser amplitude, $n_e$ is the plasma electron density, and $n_c =m_e\omega_0^2/4\pi e^2$ is the critical density for a laser with frequency $\omega_0$.

The previous Buckingham analysis has demonstrated that there cannot be any such similarity parameter in the problem we consider here. As soon as the ions are allowed to move, the electron to ion mass ratio $R$ \emph{must} appears explicitly in any list of dimensionless parameters describing the system. Two systems differing only by their mass ratio will not evolve similarly.

A deviation of the simulation dynamics from the true plasma dynamics is acceptable, as long the modifications are only quantitative. The simulation can in this case still provide important qualitative insight into the plasma evolution, which can not be obtained by any other means. However, somewhere in between the mass ratio of 1/1836 and the (positron) mass ratio of 1, a line must be crossed when even this is not the case any more.

For an unstable system with ratio $R=m_i / m_e$ between the ion and the electron mass, the unstable spectrum $\mathbb{S}=\{\mathbf{k}\in \mathbb{R}^3~/~\delta(\mathbf{k},R)>0\}$ is comprised of all the modes with wavenumbers $\mathbf{k}$ with an amplitude that grows at the exponential (positive) growth rate $\delta$. Among these modes, the most unstable mode $\mathbf{k}_m(R)$ defined by,
\begin{equation}\label{eq:k_m}
 \delta(\mathbf{k}_m,R)=\max \{\delta(\mathbf{k},R)\}_{\mathbf{k}\in \mathbb{S}}\equiv \delta_m(R),
\end{equation}
plays a peculiar role because it is the one which growth determines the outcome of the linear phase. The evolution of $\mathbf{k}_m$, as a function of $R$, may be continuous, or not. For clarity, let us consider the example studied in Sec. \ref{sec:1D}, of a one dimensional beam-plasma system. The dispersion equation in this case gives two kinds of unstable modes: the two-stream modes and the Buneman modes. Let us assume that we have plasma parameters that are such that the dominant mode $\mathbf{k}_m$ is a two-stream mode. It is possible to find a range of mass ratios $R$, for which the two-stream mode always grows fastest. In this case, $\mathbf{k}_m$ evolves continuously with $R$. A variation of the mass ratio may trigger in another case a transition from the two-stream regime to a Buneman regime where $\mathbf{k}_m$ is the wavenumber of a Buneman mode. Here, the evolution of $\mathbf{k}_m$ will be discontinuous and we will talk about an \emph{altered mode hierarchy}.

Changing the mass ratio in such a way that the mode hierarchy is altered will thus result in a different plasma evolution during the linear growth phase of the instabilities. In our 1D example, the typical size of the patterns generated in the early evolution will change abruptly by a factor $n_b/n_e$, where $n_{b,e}$ are the beam and plasma electronic densities respectively. For the 2D system considered in Sec. \ref{sec:2D}, a switch from a two-stream to a filamentation regime results in the generation of magnetized filaments instead of electrostatic stripes.

Although we focus in what follows on a specific setup, most kinds of beam-plasma systems encountered in the literature also exhibit more than one type of unstable modes  \cite{califano3,Bell2004,bretApJ2009}.

The criterion we propose is that \emph{the modified mass ratio must not alter the mode hierarchy}. Note that this is a necessary but not a sufficient condition. If the mode hierarchy is altered, then the evolution of the system should be affected as well. However, even a similar linear phase could result in a different non-linear long-term evolution prompted by a different mass ratio.

Even if thermal effects are neglected, the analysis of the full spectrum of unstable waves is involved for energetic astrophysical plasmas \cite{bretApJ2009}. The identification of the fastest-growing wave mode requires the evaluation of the full three-dimensional spectrum of wave vectors \cite{califano1,califano2,califano3,BretPRL2005}. In what follows, the proposed criterion for the mass ratio is applied to the simple and generic, yet important, system formed by a relativistic electron beam that passes through a plasma with an electronic return current. The return current initially cancels the beam current and the ion charge density cancels the total electronic one. Since some PIC simulations are still performed in a 1D geometry, we start by analyzing the 1D case before we turn to the more realistic 2D and 3D ones.

\subsection{Relativistic electron beam - 1D simulation}\label{sec:1D}
We consider a relativistic electron beam with the density $n_b$, the velocity $\mathbf{v}_b$ and the Lorentz factor $\gamma_b=(1-\mathbf{v}_b^2/c^2)^{-1/2}$, which passes through a plasma with the ion density $n_i$ and the electron density $n_e$ with $n_i=n_b+n_e$. The drift velocity $\mathbf{v}_e$ of the electrons of the background plasma fulfills $n_b\mathbf{v}_b+n_e\mathbf{v}_e=0$. For the stability analysis, we consider the response of the system to harmonic perturbations $\propto\exp(i\mathbf{k}\cdot\mathbf{r}-i\omega t)$. We assume that the particles move along the $z$-axis and we consider only wavevectors with $\mathbf{k}\parallel \mathbf{z}$. All plasma species are cold. The dispersion equation is readily expressed as the sum of the contributions by the beam, by the return current and by the ions with mass $M_i$.
\begin{equation}\label{eq:coldDim}
1=\frac{4\pi n_ie^2/M_i}{\omega^2}+\frac{4\pi n_be^2/m_e }{(\omega-kv_b)^2\gamma_b^3}+\frac{4\pi n_e e^2/m_e}{(\omega +kv_e)^2\gamma_e^3}.
\end{equation}
where $\gamma_e = (1-\mathbf{v}_e^2/c^2)^{-1/2}$ is the Lorentz factor of the return current. We introduce with $\omega_e^2=4\pi n_e e^2/m_e$ the dimensionless variables
\begin{equation}\label{eq:dimless}
\alpha=\frac{n_b}{n_e},~~Z=\frac{kv_b}{\omega_e},~~R=\frac{m_e}{M_i}.
\end{equation}
The dispersion equation expressed in these variables is
\begin{equation}\label{eq:cold}
1=\frac{R(1+\alpha)}{x^2}+\frac{\alpha }{(x-Z)^2\gamma_b^3}+\frac{1}{(x +\alpha Z)^2\gamma_e^3}.
\end{equation}
As long as $\alpha \ll 1$ the return current remains non-relativistic with $\gamma_e\sim 1$. For the strictly symmetric case with $\alpha=1$, the return current becomes relativistic with $\gamma_e=\gamma_b$. The dispersion equation (\ref{eq:cold}) defines two kinds of unstable modes \cite{BretPoPIons}. The two-stream instability is driven by the two electron beams. In the limit $\alpha\ll 1$ this instability has its maximum growth rate $\delta$ at the wavenumber
\begin{equation}\label{eq:coldTSebeam}
Z\sim 1,~~\mathrm{with}~~\delta\sim \frac{\sqrt{3}}{2^{4/3}}\frac{\alpha^{1/3}}{\gamma_b}.
\end{equation}

\begin{center}
\begin{figure}
\includegraphics[width=0.45\textwidth]{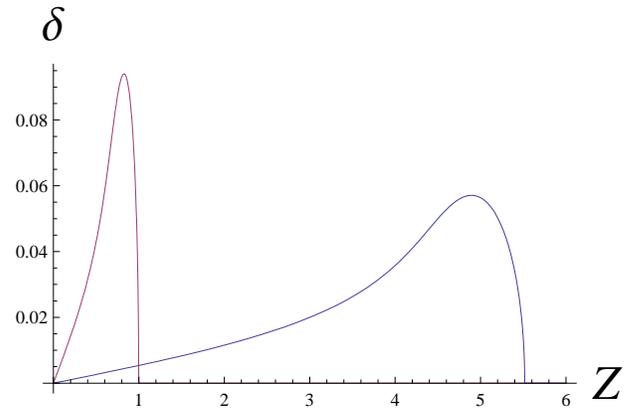}
\caption{(Color online) Growth rates of the two branches of unstable electrostatic modes derived from Eq. (\ref{eq:cold}) for $\alpha=0.2$, $R=1/1836$ and $\gamma_b=4$. Two-stream modes reach their maximum growth rate for $Z\sim 1$ and Buneman modes for $Z\sim 1/\alpha$.}
\label{fig1}
\end{figure}
\end{center}
The unstable Buneman modes arise from the interaction of the electronic return current with the ions. These additional modes grow for $\alpha \ll 1$ at the wavenumber
\begin{equation}\label{eq:coldBun}
Z\sim 1/\alpha,~~\mathrm{with}~~\delta\sim \frac{\sqrt{3}}{2^{4/3}}R^{1/3}.
\end{equation}
Figure \ref{fig1} displays the growth rate curves obtained from Eq. (\ref{eq:cold}). Both wave branches share the same $Z$-interval if $\alpha \sim 1$. Equations (\ref{eq:coldTSebeam},\ref{eq:coldBun}) show how the mode hierarchy relies explicitly on the mass ratio. Only the growth rate of the Buneman modes scales like $R^{1/3}$. Changing the mass ratio can thus change the mode hierarchy.

\begin{center}
\begin{figure}
\includegraphics[width=0.45\textwidth]{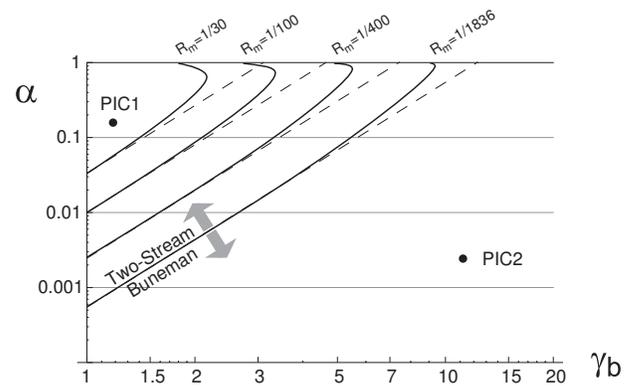}
\caption{Separatrices of the parameter space intervals dominated either by the two-stream instability or by the Buneman instability. The curves $R_m$ correspond to different mass ratios for the electrostatic 1D system considered in Section \ref{sec:1D}. Plain lines: numerical evaluation. Dashed lines: borders defined by Eq. (\ref{eq:Rm}).}
\label{fig2}
\end{figure}
\end{center}

Figure \ref{fig2} depicts the range of parameters $(\gamma_b,\alpha)$ where two-stream and Buneman modes govern the spectrum for various mass ratios $R$. The separatrix between the domain is plotted for $R=1/30$, 1/100, 1/400 and 1/1836. The Buneman modes grow faster below the curve, while the two-stream modes are dominant above. The separatrix $R_M$ between the two domains is given by
\begin{equation}\label{eq:Rm}
R_m=\frac{\alpha}{\gamma_b^3}.
\end{equation}

Let us assume that we run a 1D PIC simulation from the parameters pictured by the circle labeled ``PIC1''. For $R=1/1836$, the corresponding system lies in the two-stream region.  As we increase $R$, the growth rate of the Buneman instability increases relative to that of the two-stream instability. For sufficiently light ions, the Buneman instability can even outgrow the two-stream instability. Given some simulation parameters $(\gamma_b,\alpha)$, the largest mass ratio that leaves the mode hierarchy unchanged is readily calculated from Eq. (\ref{eq:Rm}) if $\alpha\ll 1$. One can see that a mass ratio as high as 1/30 is allowed only for weakly relativistic systems. If a PIC simulation uses the parameter values denoted by ``PIC2'', the present criterion does not restrict the mass ratio. Any value larger than $1/1836$ would be in favor of the Buneman modes, which are already governing the system.

\subsection{Relativistic electron beam - 2D and 3D simulations}\label{sec:2D}
The previous reasoning is now expanded to a 2D geometry. It is equivalent to a full 3D geometry with regard to a linearized theory, as long as the system is cold and does not have two distinct symmetry axes. An example is a beam velocity vector $\mathbf{v}_b$ that is not aligned with the magnetic field direction. Here, the $\mathbf{v}_b$ forms the sole symmetry axis and it defines one direction. A second dimension takes into account the unstable modes with wave vectors that are not parallel to $\mathbf{v}_b$. These modes compete with the two-stream and Buneman modes. One finds the Weibel (or filamentation) modes for $\mathbf{k}\perp \mathbf{v}_b$, which could play a major role in the magnetic field generation that is necessary to explain Gamma Ray Bursts \cite{Medvedev1999,Brainerd2000}. For obliquely oriented wave vectors, the so-called ``oblique modes'' are likely to govern parts of the relativistic regime \cite{BretPRL2008}.

\begin{center}
\begin{figure}
\includegraphics[width=0.45\textwidth]{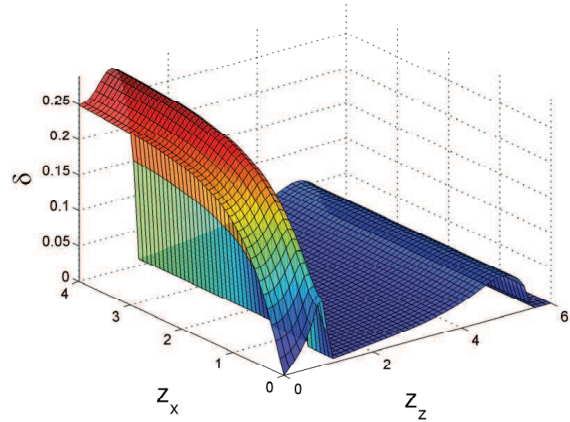}
\caption{(Color online) Growth rate map as a function of $Z_x,Z_z$. The parameters are the same as in Fig. \ref{fig2}. The beam velocity vector points along the $z$ axis.}
\label{fig3}
\end{figure}
\end{center}

The dispersion equation is more involved in 2D than in 1D, because unstable modes are generally not longitudinal (i.e. electrostatic with $\mathbf{k}\parallel \mathbf{E}$). While oblique unstable modes have been known to exist for some decades now \cite{Watson,Bludman,fainberg}, the first exact cold fluid analysis of the full unstable spectrum was only recently performed by Califano \emph{et. al.} \cite{califano3}. The dielectric tensor is computed exactly from the Maxwell's equations, the continuity equation and the Euler equation for the three species involved. We choose $\mathbf{v}_b \parallel \mathbf{z}$ and a wave vector $(k_x,0,k_z)$ in the ($x,z$) plane. The normalized wave vector $Z$ from Eq. (\ref{eq:dimless}) is now extended to two dimensions,
\begin{equation}\label{eq:2Dk}
\mathbf{Z}=\frac{\mathbf{k}v_b}{\omega_e}.
\end{equation}
The dielectric tensor has been computed symbolically using a \emph{Mathematica} Notebook designed for this purpose \cite{BretCPC}. The dispersion equation reads now
\begin{equation}\label{eq:disperT}
    \det( \mathcal{T})=0,
\end{equation}
where the tensor $ \mathcal{T}$ is specified in the Appendix. The dispersion equation is an 8th degree polynomial. Its numerical solution is straightforward. Figure \ref{fig3} shows a plot of the growth rate in terms of $\mathbf{Z}=(Z_x,Z_z)$ for the same parameters as Fig. \ref{fig1}. The Weibel or filamentation modes are characterized by $Z_z=0$, the two-stream and Buneman modes by $Z_x=0$, while the oblique modes constitute the remaining spectrum. The ridge in the growth rate map at low $Z_z$ stems from the interaction of the two electron beams, while the interaction of the ions with the bulk electrons is responsible for that at larger $Z_z$'s.

\begin{center}
\begin{figure}
\includegraphics[width=0.45\textwidth]{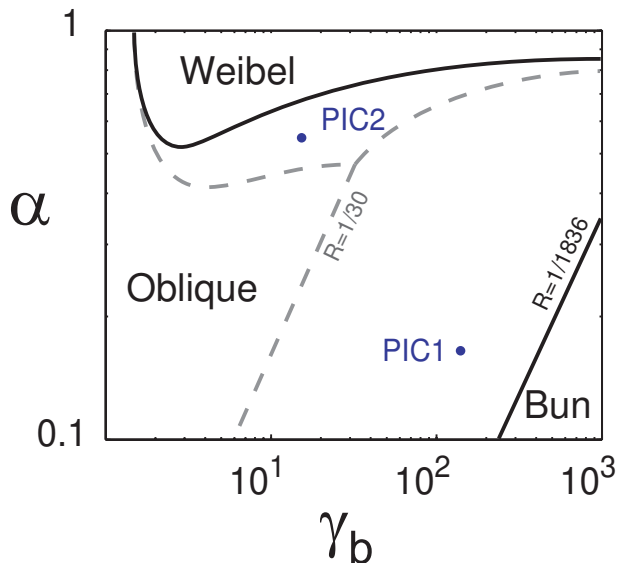}
\caption{(Color online) 2D hierarchy map in terms of $(\gamma_b,\alpha$). Plain lines: $R=1/1836$. Dashed lines: $R=1/30$. Weibel instability tend to govern the high density regime, Buneman the ultra-relativistic one, and oblique the rest of the phase space.}
\label{fig4}
\end{figure}
\end{center}

The growth rates of the Weibel modes and of the oblique modes can be estimated for immobile ions with $R=0$ and for low $\alpha$ as \cite{fainberg},
\begin{eqnarray}\label{eq:WeibObli}
\mathrm{Weibel}:~~\delta_{W}&=&\beta\sqrt{\frac{\alpha}{\gamma_b}}\nonumber\\
\mathrm{Oblique}:~~\delta_{O}&=&\frac{\sqrt{3}}{2^{4/3}}\left(\frac{\alpha}{\gamma_b}\right)^{1/3}.
\end{eqnarray}
These expressions must be corrected in a nontrivial way in the ultra-relativistic limit as $\alpha$ approaches unity. Prior to the formulation of our criterion for the mass ratio $R$, we elucidate the hierarchy map and how it evolves with $R$. Figure \ref{fig4} pictures the separatrices of the domains in the parameter space for $R=1/1836$ and for $R=1/30$. Weibel modes tend to govern the high beam density regime, and slightly expand the mildly-relativistic (around $\gamma_b=20$) part of their domain when $R$ grows. Buneman modes modes govern the lower-right corner of the graph, and being scaled like $R^{1/3}$, increase their domain as well with $R$. As a result, the domains governed by the oblique modes shrinks with a growing $R$. The two-stream modes actually never govern the system, because they are outgrown by the oblique modes as soon as $\gamma_b>1$.

\begin{center}
\begin{figure}
\includegraphics[width=0.45\textwidth]{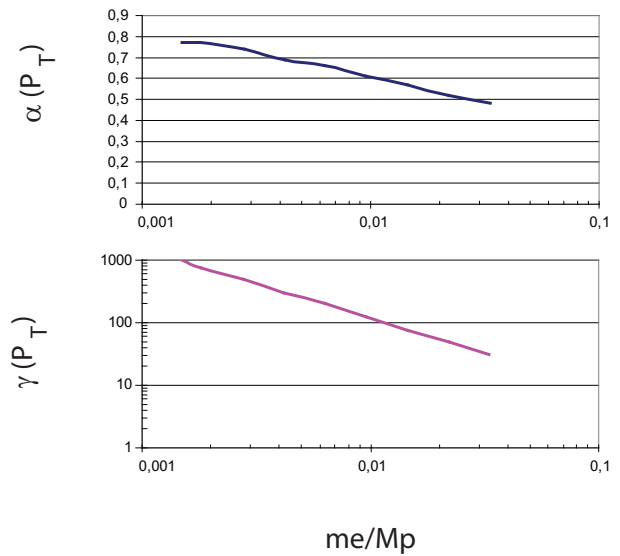}
\caption{(Color online) Coordinates of the triple point where oblique, Weibel and Buneman modes grows exactly the same rate, in terms of the electron to proton mass ratio.}
\label{fig:triple}
\end{figure}
\end{center}

The parameter space diagram reveals a triple point, at which the separatrices merge. For $R=1/30$, its coordinates are $(\gamma_b,\alpha)\sim(30,0.48)$. Figure \ref{fig:triple} shows how the triple point location evolves towards the ultra-relativistic regime $\gamma_b\gg 10^3$ as the ion mass is increased to that of a proton. This triple point is not likely to be important in astrophysical flows, since even the Lorentz factors of GRB jets do not reach such high values. However, it can become an issue in PIC simulations, where mass ratios of 30 and Lorentz factors of a few tens are not uncommon.

\begin{center}
\begin{figure}
\includegraphics[width=0.45\textwidth]{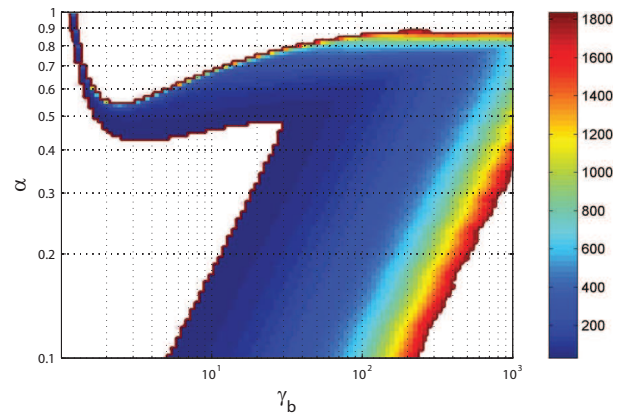}
\caption{(Color online) Smallest inverse mass ratio value $R_m^{-1}$ leaving unchanged the 2D modes hierarchy for a given parameter set $(\gamma_b,\alpha)$ for $1/1836<R_m<1/30$.
The uniform white region refers to configuration where the present criterion does not constrain the mass ratio.}
\label{fig5}
\end{figure}
\end{center}

We now compute the largest $R$ that leaves unchanged the mode hierarchy for a given parameter set $(\gamma_b,\alpha)$, and display the result on Fig. \ref{fig5} (in fact, the smallest inverse mass ratio $R_m^{-1}$ is plotted, for better clarity). If, for $R=1/1836$, a system lies in the Buneman region, then increasing $R$ will not change the dominant mode. The same holds for the systems pertaining initially to the Weibel zone. But the system represented by ``PIC1'' on Fig. \ref{fig4} remains governed by the oblique modes only up to a certain value of the mass ratio, beyond which it goes over into the Buneman domain. The same can be said for ``PIC2'': initially lying in the oblique domain, it goes over into the Weibel domain beyond a critical value of the mass ratio. Only systems already located in the oblique domain for $R=1/30$ continue to do so as we alter $R$ from $1/1836$ to $1/30$. Of course, we speak here only about the lower part of the graph that corresponds to small values of $\alpha$. The dominant mode depends through $R_m (\gamma_b,\alpha)$ critically and in a nontrivial way on both, $\alpha$ and on $\gamma_b$ in the upper part of Fig. \ref{fig5}.

We thus find a significantly extended region of the parameter space, namely, the uniform white domain on Fig. \ref{fig5}, where the criterion that the mode hierarchy be unchanged does not restrict the value of the mass ratio. For a system lying in this region, the dominant mode is the same, regardless of whether $R=1/1836$ or $1/30$. In the lower-right corner (i.e., diluted ultra-relativistic beams), the border is defined by the equality of the oblique mode (see Eq. \ref{eq:WeibObli}) with the Buneman one for $R=1/1836$,
\begin{equation}\label{eq:border1}
\alpha=R \gamma_b|_{R=1/1836}.
\end{equation}
In the lower-left corner (i.e, diluted, weakly relativistic regime), the border is determined by equating the oblique growth rate with the Buneman one, but now for $R=1/30$,
\begin{equation}\label{eq:border3}
\alpha=R \gamma_b|_{R=1/30}.
\end{equation}
The upper-border of the uniform Weibel domain is analytically more intricate. Let us just mention that the particular shape exhibited for $\gamma_b\sim 2$ arises from the Weibel growth rate which reaches a maximum around this value. Expression (\ref{eq:WeibObli}) for this quantity makes it clear that $\delta_{W}(v_b=0)=0$, while $\lim_{v_b\rightarrow c}\delta_{W}=0$. As a consequence, $\delta_{W}$ reaches a maximum for an intermediate Lorentz factor $\gamma_b=\sqrt{3}$, which is easily calculated from Eq. (\ref{eq:WeibObli}). Although this value is not exact for $\alpha$ close to unity and for $R\neq 0$, the ``Weibel optimum'' for $\gamma_b$ stays close to $\sqrt{3}$, explaining the ``bump'' at this location.

\section{Conclusion}
The importance of the electron to ion mass ratio $R = m_e / m_i$ for the realism of PIC simulations has been addressed here from an analytical point of view. Since there cannot be any rigorous similarity theory encompassing this quantity, an attempt has been made to identify a threshold $R_m$, beyond which a given simulation can no longer be trusted  to be physically accurate during the initial exponential growth phase of the instability. This initial wave growth can be addressed by a linearized theory.

Whether it be relevant for astrophysical plasmas or for inertial fusion, many systems investigated through PIC simulations give rise to the growth of waves that can be addressed by an analysis of the linear dispersion relation. The idea is therefore to find the maximum value of $R$, which leaves unchanged the hierarchy of the linearly unstable modes.  The condition we propose is necessary but not sufficient: for the system evolution to be preserved, the linear evolution and, more specifically, the type of the fastest growing mode must remain unchanged as we change $R$. However, two similar linear growth phases can eventually result in a different non-linear state.

Because the application of the criterion depends on the linear unstable spectrum, and therefore on the system under scrutiny, we have focused on the generic system formed by a relativistic electron beam passing through a plasma with return current. For a 1D simulation, the competing modes are the two-stream mode and the Buneman mode (see Fig. \ref{fig2}). The criterion of the preserved mode hierarchy does provide a value of $R_m$, if the spectrum is governed by the two-stream instability for $R=1/1836$. As a result, for example, the simulation of a 10 times diluted beam with $\gamma_b=2$ cannot be performed with ions that have a mass below $\sim 100$ times heavier than that of the electrons. For systems governed by the Buneman instability when $R=1/1836$, our criterion does not give any upper value of $R_m$.

The 2/3D case is even more interesting as more modes compete in the linear phase. Here, the Buneman, the oblique and the Weibel instabilities can dominate the linear phase, while the two-stream instability is unimportant for relativistic beam speeds. For $R=1/30$ and 1/1836, the hierarchy map is plotted on Fig. \ref{fig3} in terms of the density ratio $\alpha$ and the beam Lorentz factor $\gamma_b$. One can notice how the upper part ($\alpha\sim 1$) does not vary with $R$.  This could explain why PIC simulations of collisions between equally dense plasma shells did not show much difference as the mass ratio has been altered \cite{Spitkovsky2008}. The dominant mode is definitely the Weibel (filamentation) one in this region. Things should be different when simulating collisions of shells with a different density, like in \cite{DieckmannApJ,Bessho}, and varying the mass ratio.

The value of $R_m$ in terms of $(\gamma_b,\alpha$) is predictable at low $\alpha$, and more involved for $\alpha\sim 1$ (see Fig. \ref{fig5}). A few points can be emphasized at this junction: First, the most sensitive points are the ones located near a border between two modes for $R=1/1836$. When $R$ departs from this value, the border moves, say from mode (A) domain to mode (B) domain, so that (A) increases to the expenses of (B). If the point was initially in the (A) domain, it remains there and the criterion is not binding. But if the point was close to the border, yet in the (B) region, then a slight increase of $R$ transfers it to the (A) region. This is why on Fig. \ref{fig5}, the white region of unconstrained $R$ always borders $R_m=1/1836$.

Second, some alteration of the mode hierarchy are more dramatic than others. Figures \ref{fig4} \& \ref{fig5} show that 3 kinds of transitions can be triggered when increasing $R$: oblique to Buneman (OB - for diluted beam), oblique to Weibel (OW - high density, weakly relativistic) and Buneman to Weibel (BW - high density, ultra-relativistic). For diluted beams, the OB transition switches the wavelength of the dominant mode from $Z_z\sim 1$ to $1/\alpha$, resulting in generated structures $\alpha$ times smaller. Furthermore, oblique modes generate partially electromagnetic transverse structures whereas the electrostatic Buneman modes do not. More dramatic can be the OW transition as we now switch from a quasi-electrostatic dominant modes to an electromagnetic one. But the BW transition is by far the most powerful as the generated patterns switch from stripes (Buneman) to Filaments (Weibel).

Note however that transitions are smoother than they appear because the switch from one mode regime to another is not immediate when a border is crossed. Suppose we move from domain (A) to (B). As we approach the border, mode (A) keeps growing faster, but mode (B) grows almost as fast, until the growth rates are strictly equal right on the border. There is therefore a zone extending on both side on the line where a proper interpretation of the linear regime needs to account for the growth of (A) \emph{plus} (B), thus smoothing out the transition.

The present article is a first step towards a systematic search. The method proposed has been applied to a generic beam-plasma system, evidencing non-trivial values of $R_m$. A similar analysis can be easily conducted varying the set-up: one needs first to evaluate the growth-rate map (the counterpart of Fig. \ref{fig3}) as a function of $\mathbf{k}$ for the system under scrutiny with $R=1/1836$. The same plot is then evaluated for the desired value of $R$. If the dominant mode remains the same, then the present criterion is met.

For example, when dealing with the problem of magnetic field amplification and particles acceleration in Supernova Remnants, a typical PIC setup consists in a non-relativistic beam of \emph{protons} passing through a plasma with a guiding magnetic field \cite{Niemiec2008,Riquelme}. Due to the magnetization, unstable modes such as the Bell's ones \cite{Bell2004} enrich the spectrum, and it would be interesting to apply our criterion also to these cases.

\begin{acknowledgements}
A Bret acknowledges the financial support  of project PAI08-0182-3162 of the Consejeria de Educacion y Ciencia de la Junta de Comunidades de Castilla-La Mancha. ME Dieckmann acknowledges financial support by the Science Foundation Ireland grant 08/RFP/PHY1694 and by Vetenskapsr{\aa}det. Thanks are due to Martin Pohl and Jacek Niemiec for useful discussions.
\end{acknowledgements}

\appendix
\section{2D and 3D Tensor}
Choosing the axis $z$ for the flow direction, and $\mathbf{Z}=(Z_x,0,Z_z)$, the tensor involved in the dispersion equation (\ref{eq:disperT}) is symmetric, and reads,
\begin{equation}\label{ap:tensor}
  \mathcal{T}=\left(%
\begin{array}{ccc}
  T_{11}   &  0            &   T_{31}  \\
  0        &      T_{22}   &   0       \\
  T_{31}   &  0            &   T_{33}  \\
\end{array}
\right),
\end{equation}
where,
\begin{eqnarray}\label{ap:tensorelts}
  T_{11} &=& 1-\frac{R (1+\alpha )}{x^2}-\frac{1}{x^2}\left[\frac{Z_z^2}{\beta ^2}+\frac{\alpha }{\gamma_b}+\frac{1}{\gamma_e}\right] ,\nonumber\\
  T_{22} &=& -\frac{\left(Z_x^2+Z_z^2\right)}{x^2 \beta ^2 }
  -\frac{\gamma_b+\alpha  \gamma_e+\left(R-x^2+R \alpha \right) \gamma_b }{x^2  \gamma_b \gamma_e},\nonumber\\
  T_{33} &=& 1-\frac{R (1+\alpha )}{x^2}-\frac{\alpha }{(x-Z_z)^2 \gamma_b^3}-\frac{1}{(x+Z_z \alpha )^2 \gamma_e^3} \nonumber\\
  && -\frac{Z_x^2}{x^2} \left[\frac{1}{\beta ^2}+\frac{\alpha }{(x-Z_z)^2 \gamma_b}+\frac{\alpha ^2}{(x+Z_z \alpha )^2 \gamma_e}\right],\nonumber\\
  T_{31} &=& \frac{Z_x}{x^2} \left[\frac{Z_z}{\beta ^2}+ \frac{\alpha }{x \gamma_e+Z_z \alpha
   \gamma_e}+\frac{\alpha }{(Z_z-x) \gamma_b}\right].\nonumber
\end{eqnarray}

%

\end{document}